\documentclass[twocolumn,english,aps,pra,showpacs,superscriptaddress]{revtex4}
\usepackage{mathptmx}
\usepackage[T1]{fontenc}
\usepackage[latin9]{inputenc}
\usepackage{color}
\usepackage{amstext}
\usepackage{amssymb}
\usepackage{graphicx}
\usepackage{esint}
\PassOptionsToPackage{normalem}{ulem}
\usepackage{ulem}

\makeatletter
\@ifundefined{textcolor}{}
{%
 \definecolor{BLACK}{gray}{0}
 \definecolor{WHITE}{gray}{1}
 \definecolor{RED}{rgb}{1,0,0}
 \definecolor{GREEN}{rgb}{0,1,0}
 \definecolor{BLUE}{rgb}{0,0,1}
 \definecolor{CYAN}{cmyk}{1,0,0,0}
 \definecolor{MAGENTA}{cmyk}{0,1,0,0}
 \definecolor{YELLOW}{cmyk}{0,0,1,0}
 }

\renewcommand{\vec}{\mathbf}
\usepackage[amssymb]{SIunits}

\makeatother

\usepackage{babel}
\begin{document}

\title{Phonon-induced entanglement dynamics of two donor-based charge quantum
bits}

\author{F. Lastra}

\affiliation{Facultad de Física, Pontificia Universidad Católica de Chile, Casilla
306, Santiago 22, Chile}

\affiliation{Departamento de Física \& Center for the Development of Nanoscience
and Nanotechnology, Universidad de Santiago de Chile, Casilla 307,
Santiago, Chile}

\author{S.A. Reyes }

\affiliation{Facultad de Física, Pontificia Universidad Católica de Chile, Casilla
306, Santiago 22, Chile}

\author{S. Wallentowitz}

\affiliation{Facultad de Física, Pontificia Universidad Católica de Chile, Casilla
306, Santiago 22, Chile}
\begin{abstract}
The entanglement dynamics of a pair of donor-based charge qubits is
obtained in analytical form. The disentanglement is induced by off
resonant scattering of acoustical phonons in the semiconductor host.
According to our results a rather unusual recovery of entanglement
occurs that depends on the geometrical configuration of the qubits.
In addition, for large times a non-vanishing stationary entanglement
is predicted. For the cases of one and two initial excitations a simple
kinetic interpretation allows for an adequate analysis of the observed
dynamics. Our results also reveal a direct relation between the disentanglement
rate and the inter-donor decoherence rates. 
\end{abstract}

\date{August 11, 2011}


\maketitle

\section{Introduction}

Quantum information processing promises highly efficient solutions
to cryptographic problems and exhaustive database search, that outperform
the best known algorithms with classical computers \cite{niel}. Such
quantum algorithms rely on the capability to process correlations
among quantum subsystems. The quantum part of these correlations is
denoted as entanglement. Most commonly, the subsystems are identified
as two-level systems, representing a quantum counterpart of classical
bits, called quantum bits (qubits). Among the well known applications
of quantum information are quantum teleportation \cite{ben}, superdense
coding \cite{ben1}, and secure distribution of cryptographic keys
\cite{ekr}.

Entanglement, being the key ingredient of such applications, at the
same time is highly fragile and can be easily deteriorated during
state preparation, addressing and control of individual qubits, and
final readout. Hence, it is necessary to encode quantum information
in physical systems, where entanglement is protected or can be preserved
in a robust way from ambient effects. Prominent examples of physical
implementations are trapped ions \cite{lei}, nuclear magnetic resonance
\cite{van}, atoms in cavities \cite{rai}, quantum dots \cite{han},
semiconductor impurities \cite{kan}, superconducting qubits \cite{dic},
and impurities in diamond \cite{gae}.

Among the solid-state implementations \cite{pro,ami,chi} impurities
embedded in a semiconductor substrate offer the advantage of comparably
easy scaling and production due to highly developed fabrication techniques.
The encoding of qubits in the charge degrees of freedom of pairs of
donor sites allows for the addressing of individual qubits by metallic
gates \cite{hol,bel}. Tunneling of the electron between donor sites
together with Coulomb repulsion of electrons bound in neighboring
qubits have been shown to allow for the realization of a CNOT gate
\cite{hol,tsu,ste}. Such a gate is the elemental building block of
any quantum algorithm. Typical coherence times of such systems are
limited by phonon scattering to the order of $1\pico\second$ \cite{zhao,non-m,las2}.
Therefore, the state preparation process must be faster and one may
expect that entanglement is rapidly lost on this time scale. 

In this work we show that to a large extent the entanglement survives
beyond this time scale and therefore suffers further degradation only
from other sources of decoherence, i.e. charge fluctuations on the
control electrodes. The disentanglement dynamics is obtained in analytical
form showing non-Markovian features, similar to the decoherence dynamics
of a single qubit \cite{eck,non-m,las2}. Furthermore, it is shown
that the disentanglement rate is directly related to inter-donor decoherence
rates for the cases of one and two initial excitations. The structure
of this rate can be explained by a simple kinetic interpretation that
allows for the determination of the disentanglement rate from the
geometry of the constituent donor sites.

The paper is organized as follows: In Sec. \ref{sec:Electronic-Dynamics}
the dynamics of a general $N$ qubit system subject to off-resonant
scattering of acoustical phonons is derived. Using these results the
dynamics of entanglement between two qubits is analytically obtained
in Sec. III for the cases of one and two excitations. Finally, in
Sec. IV we present a summary and conclusions.

\section{Phonon-induced dephasing dynamics of qubits\label{sec:Electronic-Dynamics}}

\begin{figure}
\begin{centering}
\includegraphics[width=1\columnwidth]{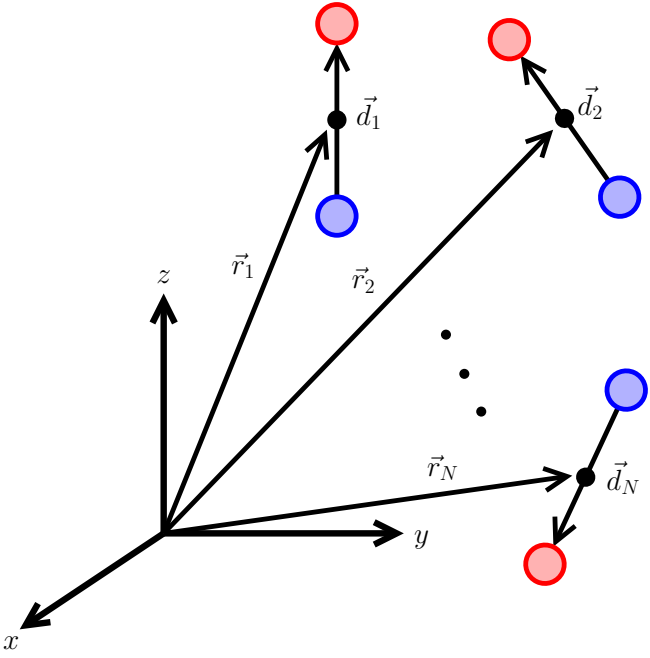}
\par\end{centering}

\caption{\label{fig:qubits}Geometrical setup of $N$ qubits formed by $2N$
donor sites in a semiconductor material. Blue and red donor sites
correspond to qubit states $m_{b}=-\frac{1}{2}$ and $m_{b}=+\frac{1}{2}$,
respectively, where $b=1,\ldots,N$ denotes the qubit. Each qubit
$b$ has an inter-site distance $\vec{d}_{b}$ pointing from site
$m_{b}=-\frac{1}{2}$ to $m_{b}=+\frac{1}{2}$, and a position vector
$\vec{r}_{b}$ pointing towards the center of the qubit. }
\end{figure}

In this chapter we deduce the dynamics of the reduced density operator
of the qubit system, which is induced by off-resonant scattering of
acoustical phonons within the semiconductor material. It is assumed
here that $2N$ donor sites are present that conform $N$ qubits,
see Fig. \ref{fig:qubits}. The results will later be specialized
to the case of $N=2$ qubits. The dynamics of a single qubit conformed
by two donor sites has been derived recently by us in Ref. \cite{non-m}.
The present work goes beyond the case of two donors, which implies
a more complex dynamics together with the possibility of studying
the entanglement between pairs of qubits.

We consider the situation where a single electron is confined to each
pair of donors, conforming the $N$ qubits being centered at the positions
$\vec{r}_{b}$, where $b=1,\ldots,N$ labels the qubit under consideration.
In addition, we assume that the distances between these qubits are
much larger than the inter-donor vectors $\vec{d}_{b}$ within the
qubits, i.e. $|\vec{r}_{b}-\vec{r}_{b^{\prime}}|\gg|\vec{d}_{b^{\prime\prime}}|$.
Under these circumstances the tunneling of electrons between different
qubits can be safely neglected. Considering the interaction of the
qubit charges with acoustical phonons in the semiconductor substrate
at room temperature or below, off-resonant phonon scattering is the
main source of decoherence in the electron dynamics \cite{non-m}.
This type of electron-phonon interaction is accurately described by
the spin-boson Hamiltonian

\begin{eqnarray}
\hat{H} & = & \hbar\sum_{b}\left(\omega_{b}\hat{S}_{b,z}+\Delta_{b}\hat{S}_{b,x}\right)+\sum_{\vec{k}}\hbar\omega_{k}\hat{a}_{\vec{k}}^{\dagger}\hat{a}_{\vec{k}}\nonumber \\
 &  & +\hbar\sum_{b}\hat{S}_{b,z}\sum_{\vec{k}}\left(g_{b,\vec{k}}\hat{a}_{\vec{k}}^{\dagger}+g_{b,\vec{k}}^{\ast}\hat{a}_{\vec{k}}\right),\label{eq:hamiltonian}
\end{eqnarray}
where the individual interaction rate of the $m_{b}$th donor site
($m_{b}=\pm1/2$) within qubit $b$ reads 
\begin{equation}
g_{b,\vec{k}}=\frac{D}{\hbar s}\sqrt{\frac{2\hbar\omega_{k}}{M_{0}}}\sum_{m_{b}=\pm1/2}\frac{m_{b}e^{-i\vec{k}\cdot(\vec{r}_{b}+m_{b}\vec{d}_{b})}}{\left[1+\left(\frac{ka_{b,m_{b}}}{2}\right)^{2}\right]^{2}}.
\end{equation}
Here $\hat{a}_{\vec{k}}$ are the annihilation operators of longitudinal
acoustical phonon with dispersion relation $\omega_{k}=sk$, $s$
being the sound speed. Moreover, $M_{0}$ and $D$ are the mass within
a unit cell of the semiconductor and the deformation constant, respectively.
Each donor site is described by a $s$-wave ground-state with Bohr
radius $a_{b,m_{b}}$ and the electronic transition of each qubit
$b$ is generated by the pseudo spin-$\frac{1}{2}$ operator $\hat{\vec{S}}_{b}$,
where 
\begin{equation}
\hat{S}_{b,z}=\frac{1}{2}\left(\biggl|\vec{r}_{b}+\frac{\vec{d}_{b}}{2}\biggr\rangle\biggl\langle\vec{r}_{b}+\frac{\vec{d}_{b}}{2}\biggr|-\biggl|\vec{r}_{b}-\frac{\vec{d}_{b}}{2}\biggr\rangle\biggl\langle\vec{r}_{b}-\frac{\vec{d}_{b}}{2}\biggr|\right),
\end{equation}
with $|\vec{r}_{b}\pm\vec{d}_{b}/2\rangle$ being the states with
the electron being localized at the corresponding donor site. In what
follows we assume that during the free evolution of the system, tunneling
is inhibited by either an applied potential barrier between the donor
sites, or due to a strong bias between the qubit levels, $|\omega_{b}|\gg|\Delta_{b}|$,
which can be provided for by the application of a DC electric field. 

Following the same steps as in Ref. \cite{non-m}, the Hamiltonian
Eq.(\ref{eq:hamiltonian}) can be diagonalized, to obtain the complete
set of eigenstates as displaced number states 
\begin{equation}
|E_{\{m_{b}\},\{N_{\vec{k}}\}}\rangle=|\{m_{b}\}\rangle\otimes\hat{D}^{\dagger}\left(\left\{ \alpha_{\{m_{b}\},\vec{k}}\right\} \right)|\{N_{\vec{k}}\}\rangle,\label{eq:eigenstates}
\end{equation}
with eigenenergies
\begin{equation}
E_{\{m_{b}\},\{N_{\vec{k}}\}}=\hbar\sum_{b}\omega_{b}m_{b}+\sum_{\vec{k}}\hbar\omega_{k}N_{\vec{k}}.\label{eq:eigenenergies}
\end{equation}
Here the multi-mode displacement operator reads
\begin{equation}
\hat{D}\left(\left\{ \alpha_{\vec{k}}\right\} \right)=\exp\left[\sum_{\vec{k}}\left(\alpha_{\vec{k}}\hat{a}_{\vec{k}}^{\dagger}-\alpha_{\vec{k}}^{\ast}\hat{a}_{\vec{k}}\right)\right],
\end{equation}
and $|\{N_{\vec{k}}\}\rangle=\prod_{\vec{k}}|N_{\vec{k}}\rangle$
are multi-mode number states of the acoustic phonons and the displacement
amplitude reads 
\begin{equation}
\alpha_{\{m_{b}\},\vec{k}}=\sum_{b}m_{b}\alpha_{b,\vec{k}},
\end{equation}
where $\alpha_{b,\vec{k}}=g_{b,\vec{k}}/\omega_{k}$. Furthermore,
the quantum state of the qubits is encoded in the register state 
\begin{equation}
|\{m_{b}\}\rangle=|m_{1}\rangle\otimes|m_{2}\rangle\otimes\cdots\otimes|m_{N}\rangle,
\end{equation}
where each of the qubits can be in states $|m_{b}=\pm\frac{1}{2}\rangle$
($b=1,\ldots,N$).

Given the eigenstates (\ref{eq:eigenstates}) and eigenenergies (\ref{eq:eigenenergies}),
the general solution of the reduced density operator of the qubits,
i.e. traced over the phonons, reads 
\begin{eqnarray}
\hat{\varrho}(t) & = & \sum_{\{m_{b}\},\{s_{b}\}}|\{m_{b}\}\rangle\langle\{s_{b}\}|\sum_{\{N_{\vec{k}}\}}\sum_{\{N_{\vec{k}}^{\prime}\}}\varrho_{\{m_{b}\},\{N_{\lambda}\};\{s_{b}\},\{N_{\lambda}^{\prime}\}}\nonumber \\
 &  & \quad\times\langle\{N_{\vec{k}}^{\prime}\}|\hat{D}\left(\left\{ \alpha_{\{s_{b}\},\vec{k}}\right\} \right)\hat{D}^{\dagger}\left(\left\{ \alpha_{\{m_{b}\},\vec{k}}\right\} \right)|\{N_{\vec{k}}\}\rangle\nonumber \\
 &  & \quad\times\exp\left[-\frac{it}{\hbar}\left(E_{\{m_{b}\},\{N_{\vec{k}}\}}-E_{\{s_{b}\},\{N_{\vec{k}}^{\prime}\}}\right)\right].\label{eq:rho-reduced}
\end{eqnarray}
Here the matrix elements of the initial density operator of the complete
electron-phonon system in the basis of the energy eigenstates (\ref{eq:eigenstates})
are
\begin{equation}
\varrho_{\{m_{b}\},\{N_{\vec{k}}\};\{s_{b}\},\{N_{\vec{k}}^{\prime}\}}=\langle E_{\{m_{b}\},\{N_{\vec{k}}\}}|\hat{\varrho}(0)|E_{\{s_{b}\},\{N_{\vec{k}}^{\prime}\}}\rangle.\label{eq:initial-complete}
\end{equation}
These density matrix elements are determined by the state-preparation
process that is utilized to set up the initial entanglement between
the qubits. A generic state-preparation process can be described as
follows:

We assume that initially the system is at low enough temperature,
$k_{{\rm B}}T\ll\hbar\omega_{b}$, such that it has relaxed completely
into a state where all the qubits are in their lowest energy states,\textcolor{green}{{}
}$m_{b}=-1/2$ ($b=1,\ldots,N$), and coexist in thermal equilibrium
with the phonons in the substrate. Starting from this state the system
undergoes a state preparation process in which the qubits can be coherently
transferred into a superposition of ground and excited states, without
affecting the quantum state of the phonons. The probability amplitudes
to transfer from the initial qubit state $\{-\frac{1}{2},-\frac{1}{2},\ldots,-\frac{1}{2}\}$
to the qubit states $\{m_{b}\}$, shall be denoted by $\psi_{\{m_{b}\}}$.
 The corresponding transition can be generated by the application
of the operator
\begin{equation}
\hat{S}_{+,\{m_{b}\}}=\Pi_{b}\left[\delta_{m_{b},\frac{1}{2}}\hat{S}_{+,b}+\left(1-\delta_{m_{b},\frac{1}{2}}\right)\hat{I}_{b}\right],
\end{equation}
where $\hat{I}_{b}$ is the identity operator in the Hilbert space
of qubit $b$.\textcolor{green}{{} }Thus, the matrix elements of the
prepared initial state (\ref{eq:initial-complete}) become
\begin{eqnarray}
\varrho_{\{m_{b}\},\{N_{\vec{k}}\};\{s_{b}\},\{N_{\vec{k}}^{\prime}\}}=\qquad\qquad\qquad\qquad\qquad\qquad\qquad\label{eq:psi-e-def-1-1}\\
\psi_{\{m_{b}\}}\psi_{\{s_{b}\}}^{\ast}\langle E_{\{m_{b}\},\{N_{\vec{k}}\}}|\hat{S}_{+,\{m_{b}\}}\hat{\varrho}_{T}\hat{S}_{+,\{s_{b}\}}^{\dagger}|E_{\{s_{b}\},\{N_{\vec{k}}^{\prime}\}}\rangle,\nonumber 
\end{eqnarray}
where 
\begin{equation}
\hat{\varrho}_{T}=\sum_{\{N_{\vec{k}}\}}P_{\{N_{\vec{k}}\}}|E_{\{-1/2,-1/2,...\},\{N_{\vec{k}}\}}\rangle\langle E_{\{-1/2,-1/2,...\},\{N_{\vec{k}}\}}|\label{eq:thermal-state}
\end{equation}
is the initial thermal state with the phonon statistics 
\begin{equation}
P_{\{N_{\vec{k}}\}}=Z^{-1}\exp\left(-\sum_{\vec{k}}\beta_{k}N_{\vec{k}}\right),\label{eq:thermal-statistics}
\end{equation}
with $Z$ satisfying $\sum_{\{N_{\vec{k}}\}}P_{\{N_{\vec{k}}\}}=1$,
and $\beta_{k}=\hbar\omega_{k}/(k_{{\rm B}}T)$.

We note that the corresponding prepared initial reduced density operator
of the qubits is pure, i.e. is of the form $|\psi(0)\rangle\langle\psi(0)|$
with the qubit state being the sought superposition 
\begin{equation}
|\psi(0)\rangle=\sum_{\{m_{b}\}}\psi_{\{m_{b}\}}|\{m_{b}\}\rangle.
\end{equation}
Thus, the presence of the phonons does not prevent a coherent preparation
of the initial qubit state. However, the initial state preparation
takes the electron-phonon system out of thermal equilibrium so that
the decoherence of the quantum state of the qubits cannot be treated
by the usual method of spectral functions \cite{leggett}.

The generic state preparation, as described above, can be implemented
for example by switching the gate voltages so that the on-site energies
within a qubit cross each other\textcolor{green}{{} }inducing Landau-Zener
transitions. At the end of the process, the system will be in a coherent
superposition of all possible states of the system depending on how
fast its two-level components where driven across the level crossing.
Alternatively, it may be implemented by time controlled tunneling
and employing the Coulomb repulsion between neighboring qubits to
generate entangled qubit states, as proposed for implementing CNOT
gates \cite{hol,tsu,ste}. Furthermore, it could also be implemented
by $\tera\hertz$ Raman transitions between the donor sites \cite{aba}. 

We note that the matrix elements of the initial \emph{reduced} electronic
density operator are obtained from Eq. (\ref{eq:rho-reduced}) as
\begin{eqnarray}
\langle\{m_{b}\}|\hat{\varrho}_{S}(0)|\{s_{b}\}\rangle & = & \sum_{\{N_{\vec{k}}\}}\sum_{\{N_{\vec{k}}^{\prime}\}}\varrho_{\{m_{b}\},\{N_{\vec{k}}\};\{s_{b}\},\{N_{\vec{k}}^{\prime}\}}\nonumber \\
 &  & \qquad\times f_{\{s_{b}\},\{N_{\vec{k}}^{\prime}\};\{m_{b}\},\{N_{\vec{k}}\}},\label{eq:initial-reduced}
\end{eqnarray}
with the non-diagonal elements containing the Franck--Condon type
transition amplitudes 
\begin{eqnarray}
f_{\{s_{b}\},\{N_{\vec{k}}^{\prime}\};\{m_{b}\},\{N_{\vec{k}}\}}=\qquad\qquad\qquad\qquad\qquad\qquad\nonumber \\
\langle\{N_{\vec{k}}^{\prime}\}|\hat{D}\left(\left\{ \alpha_{\{s_{b}\},\vec{k}}\right\} \right)\hat{D}^{\dagger}\left(\left\{ \alpha_{\{m_{b}\},\vec{k}}\right\} \right)|\{N_{\vec{k}}\}\rangle.
\end{eqnarray}
These factors are overlap integrals of two displaced phonon number
states with displacements $\alpha_{\{s_{b}\},\vec{k}}$ and $\alpha_{\{m_{b}\},\vec{k}}$,
respectively. Their\textcolor{green}{{} }presence\textcolor{green}{{}
}is due to the fact that the initial density matrix of the complete
electron-phonon system (\ref{eq:initial-complete}) is in the basis
of the\textcolor{green}{{} }energy eigenstates (\ref{eq:eigenstates}),
whereas the matrix elements of Eq. (\ref{eq:initial-reduced}) are
in the basis of the product states $|\{m_{b}\}\rangle\otimes|\{N_{\vec{k}}\}\rangle$
that differ by the displacement of the phonons. 

Consistent with a dephasing model, the diagonal elements follow from
Eq. (\ref{eq:rho-reduced}) as invariants: 
\begin{equation}
\langle\{m_{b}\}|\hat{\varrho}_{S}(t)|\{m_{b}\}\rangle=\langle\{m_{b}\}|\hat{\varrho}_{S}(0)|\{m_{b}\}\rangle.
\end{equation}
However, this dephasing --- being induced by off-resonant scattering
of acoustical phonons --- modifies the time evolution of the off-diagonal
density matrix elements as 
\begin{eqnarray}
\langle\{m_{b}\}|\hat{\varrho}_{S}(t)|\{s_{b}\}\rangle & = & e^{-i\sum_{b}(m_{b}-s_{b})\omega_{b}t}\label{eq:rho-ge}\\
 & \times & \sum_{\{N_{\vec{k}}\}}\sum_{\{N_{\vec{k}}^{\prime}\}}\varrho_{\{m_{b}\},\{N_{\vec{k}}\};\{s_{b}\},\{N_{\vec{k}}^{\prime}\}}\nonumber \\
 & \times & f_{\{s_{b}\},\{N_{\vec{k}}^{\prime}\};\{m_{b}\},\{N_{\vec{k}}\}}e^{-i\sum_{\vec{k}}\omega_{k}(N_{\vec{k}}-N_{\vec{k}}^{\prime})t}.\nonumber 
\end{eqnarray}
It can be observed in Eq. (\ref{eq:rho-ge}) that apart from the free
oscillation with angular frequency $\omega_{0}$, a dephasing is induced
by\textcolor{green}{{} }differing phonon numbers in combination with
the presence of the Franck--Condon factor.\textcolor{green}{\sout{ }}

To further evaluate the dephasing of the off-diagonal density matrix
elements of the $N$ qubits, we insert the initial complete density
matrix elements (\ref{eq:initial-complete}) together with Eq. (\ref{eq:thermal-state})
into Eq. (\ref{eq:rho-ge}). From this we obtain\begin{widetext}
\begin{eqnarray}
\langle\{m_{b}\}|\hat{\varrho}_{S}(t)|\{s_{b}\}\rangle & = & \psi_{\{m_{b}\}}\psi_{\{s_{b}\}}^{\ast}e^{-i\sum_{b}(m_{b}-s_{b})\omega_{b}t}\sum_{\{M_{\vec{k}}\}}P_{\{M_{\vec{k}}\}}\sum_{\{N_{\vec{k}}\}}\sum_{\{N_{\vec{k}}^{\prime}\}}e^{-i\sum_{\vec{k}}\omega_{k}(N_{\vec{k}}-N_{\vec{k}}^{\prime})t}\nonumber \\
 & \times & \times f_{\{s_{b}\},\{N_{\vec{k}}^{\prime}\};\{-\frac{1}{2},\ldots,-\frac{1}{2}\},\{M_{\vec{k}}\}}^{\ast}f_{\{s_{b}\},\{N_{\vec{k}}^{\prime}\};\{m_{b}\},\{N_{\vec{k}}\}}f_{\{m_{b}\},\{N_{\vec{k}}\};\{-\frac{1}{2},\ldots,-\frac{1}{2}\},\{M_{\vec{k}}\}}.\label{eq:rho-ge-1-2}
\end{eqnarray}
Employing the thermal phonon statistics (\ref{eq:thermal-statistics})
the sum over the phonon numbers in Eq. (\ref{eq:rho-ge-1-2}) can
be rewritten as a trace, which leaves us with
\begin{eqnarray}
\langle\{m_{b}\}|\hat{\varrho}_{S}(t)|\{s_{b}\}\rangle & = & \psi_{\{m_{b}\}}\psi_{\{s_{b}\}}^{\ast}e^{-i\sum_{b}(m_{b}-s_{b})\omega_{b}t}Z^{-1}{\rm Tr}\left[\hat{D}(\{\alpha_{\{-\frac{1}{2},\ldots,-\frac{1}{2}\},\vec{k}}\})\hat{D}^{\dagger}(\{\alpha_{\{s_{b}\},\vec{k}}\})\right.\nonumber \\
 &  & \times\left.\hat{D}(\{\alpha_{\{s_{b}\},\vec{k}}(t)\})\hat{D}^{\dagger}(\{\alpha_{\{m_{b}\},\vec{k}}(t)\})\hat{D}(\{\alpha_{\{m_{b}\},\vec{k}}\})\hat{D}^{\dagger}(\{\alpha_{\{-\frac{1}{2},\ldots,-\frac{1}{2}\},\vec{k}}\})e^{-\sum_{\vec{k}}\beta_{k}\hat{N}_{\vec{k}}}\right],\label{eq:rho-ge-1-1}
\end{eqnarray}
\end{widetext}where we defined the time-dependent phonon displacement
amplitude $\alpha_{\{m_{b}\},\vec{k}}(t)=\alpha_{\{m_{b}\},\vec{k}}\exp(-i\omega_{k}t)$.
The displacement operators in Eq. (\ref{eq:rho-ge-1-1}) can be combined
to obtain 
\begin{eqnarray}
\langle\{m_{b}\}|\hat{\varrho}_{S}(t)|\{s_{b}\}\rangle=\psi_{\{m_{b}\}}\psi_{\{s_{b}\}}^{*}e^{-i\sum_{b}(m_{b}-s_{b})\omega_{b}t+i\Delta_{\{m_{b}\},\{s_{b}\}}(t)}\nonumber \\
\times Z^{-1}{\rm Tr}\left[\hat{D}\left(\delta_{\{m_{b}\},\{s_{b}\},\vec{k}}\left(1-e^{-i\omega_{k}t}\right)\right)e^{-\sum_{\vec{k}}\beta_{k}\hat{N}_{\vec{k}}}\right],\qquad\label{eq:rho-offdiag}
\end{eqnarray}
 with the time dependent phase being 

\begin{equation}
\Delta_{\{m_{b}\},\{s_{b}\}}(t)=\Im\sum_{\vec{k}}\bar{\alpha}_{\{m_{b}\},\{s_{b}\},\vec{k}}\delta_{\{m_{b}\},\{s_{b}\},\vec{k}}^{\ast}\left(e^{i\omega_{k}t}-1\right)
\end{equation}
Here we have defined sum and difference displacements: 
\begin{eqnarray}
\bar{\alpha}_{\{m_{b}\},\{s_{b}\},\vec{k}} & = & \alpha_{\{m_{b}\},\vec{k}}+\alpha_{\{s_{b}\},\vec{k}}-2\alpha_{\{-\frac{1}{2},\ldots,-\frac{1}{2}\},\vec{k}},\\
\delta_{\{m_{b}\},\{s_{b}\},\vec{k}} & = & \alpha_{\{m_{b}\},\vec{k}}-\alpha_{\{s_{b}\},\vec{k}}.
\end{eqnarray}

The trace in Eq. (\ref{eq:rho-offdiag}) represents a thermal average
that can be evaluated in phase space to obtain 
\begin{eqnarray}
\langle\{m_{b}\}|\hat{\varrho}_{S}(t)|\{s_{b}\}\rangle & = & \psi_{\{m_{b}\}}\psi_{\{s_{b}\}}^{*}e^{-i\sum_{b}(m_{b}-s_{b})\omega_{b}t+i\Delta_{\{m_{b}\},\{s_{b}\}}(t)}\nonumber \\
 &  & \times\exp\left\{ -\int_{0}^{t}dt^{\prime}\Gamma_{\{m_{b}\},\{s_{b}\}}(t^{\prime})\right\} ,\label{eq:rho-offdiag2}
\end{eqnarray}
where the decoherence rate of the qubit state is defined as 
\begin{eqnarray}
\Gamma_{\{m_{b}\},\{s_{b}\}}(t) & = & \sum_{b,b'}(m_{b}-s_{b})(m_{b'}-s_{b'})\gamma_{b,b'}(t).\label{eq:decrate}
\end{eqnarray}
Whereas this rate depends on the quantum numbers of the density matrix
element under consideration, the relation of pairs of bits is governed
by the inter-bit decorrelation rate, given by 
\begin{eqnarray}
\gamma_{b,b'}(t) & =4 & \sum_{m_{b}}\sum_{s_{b^{\prime}}}m_{b}s_{b^{\prime}}\gamma\left(t;a_{b,m_{b}},a_{b',s_{b'}},l_{b,m_{b};b',s_{b'}}\right).\label{eq:inter-bit-rate}
\end{eqnarray}
This rate in turn depends via the inter-donor decoherence rate $\gamma(t;a,a^{\prime},l)$
on the Bohr radii of the four donor sites of the two qubits and on
the six possible distances between these four donors, 
\begin{equation}
l_{b,m_{b};b',s_{b'}}=\left|(\vec{r}_{b}+m_{b}\vec{d}_{b})-(\vec{r}_{b^{\prime}}+s_{b'}\vec{d}_{b'})\right|.\label{eq:inter-bit-distance}
\end{equation}
Moreover, the inter-donor decoherence rate is obtained from Eqs. (\ref{eq:rho-offdiag})
-- (\ref{eq:inter-bit-distance}) as\begin{widetext}
\begin{equation}
\gamma(t;a,a^{\prime},l)=\Gamma_{T}\left[\left(\frac{aa_{{\rm B}}}{a^{2}-a^{\prime2}}\right)^{2}\sum_{\sigma=\pm1}\sigma\left(\frac{|l-\sigma st|}{l}+\frac{a}{2l}\frac{a^{2}-5a^{\prime2}}{a^{2}-a^{\prime2}}\right)e^{-2|l-\sigma st|/a}+(a\leftrightarrow a^{\prime})\right].\label{eq:inter-donor-rate}
\end{equation}
\end{widetext}In this expression the temperature dependent rate reads
$\Gamma_{T}=\omega_{{\rm B}}(T/T_{{\rm B}})$, where the convenient
temperature scale is chosen as
\begin{equation}
k_{{\rm B}}T_{{\rm B}}=N_{{\rm B}}M_{0}s^{2}\left(\frac{\hbar\omega_{{\rm B}}}{D}\right)^{2}.
\end{equation}
Here $\omega_{{\rm B}}=2\pi s/a_{{\rm B}}$ with $a_{{\rm B}}$ being
the average Bohr radius of all donor sites and $N_{{\rm B}}$ is the
number of unit cells within the average Bohr volume $a_{{\rm B}}^{3}$.

\begin{figure}
\begin{centering}
\includegraphics[width=1\columnwidth]{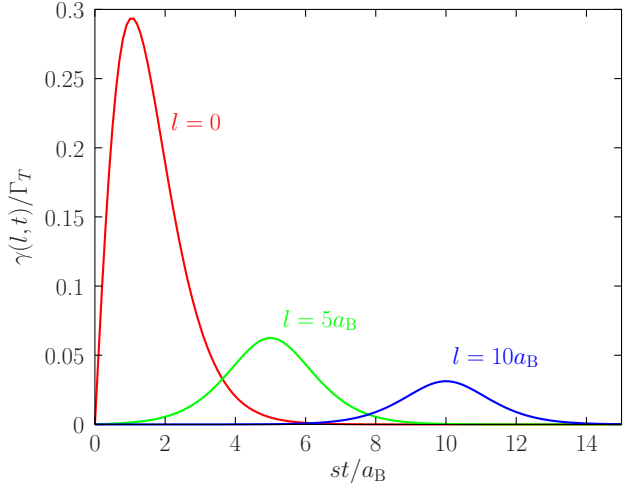}
\par\end{centering}

\caption{Dimensionless and temperature independent inter-site decoherence rate
$\gamma(l;t)/\Gamma_{T}$ as a function of the traveled distance of
a phonon in units of the Bohr radius, $st/a_{{\rm B}}$. \label{fig:Dimensionless-and-temperature}}
\end{figure}

In the limit of identical donor sites, $a_{b,m_{b}}\to a_{{\rm B}}$
($b=1,\ldots,N$), the inter-donor decoherence rate (\ref{eq:inter-donor-rate})
approaches the form $\gamma(t;a,a^{\prime},l)\to\gamma(t;l)$ with\begin{widetext}
\begin{equation}
\gamma(t;l)=\Gamma_{T}\frac{a_{B}}{l}\sum_{\sigma=\pm1}\sigma\left[\frac{1}{6}\left(\frac{|l-\sigma st|}{a_{{\rm B}}}\right)^{3}+\frac{1}{2}\left(\frac{|l-\sigma st|}{a_{{\rm B}}}\right)^{2}+\frac{5}{8}\left(\frac{|l-\sigma st|}{a_{{\rm B}}}\right)+\frac{5}{16}\right]e^{-2|l-\sigma st|/a_{{\rm B}}}.
\end{equation}
\end{widetext}We note that in the limit of vanishing distance between
donor sites, $l\to0$, this function becomes 
\begin{equation}
\gamma(t;0)=\Gamma_{T}\left[\frac{2}{3}\left(\frac{st}{a_{{\rm B}}}\right)^{3}+\left(\frac{st}{a_{{\rm B}}}\right)^{2}+\frac{1}{2}\left(\frac{st}{a_{{\rm B}}}\right)\right]e^{-2st/a_{{\rm B}}}.
\end{equation}
This is a function peaked at $t\sim a_{{\rm B}}/s$, i.e. at the time
a phonon needs to travel the distance of one Bohr radius, see Fig.
\ref{fig:Dimensionless-and-temperature} (red curve). Different from
this special case ($l\to0$) for $l>0$ the inter-donor decoherence
rate is peaked at the time $t\sim l/s$ that is required for a phonon
to travel the distance $l$, see Fig. \ref{fig:Dimensionless-and-temperature}
(green and blue curves).

\section{Dynamics of entanglement between two qubits}

In the following we will discuss the special case of two qubits being
present in the semiconductor system, i.e. $N=2$. In this case the
state of the qubits $|\{m_{b}\}\rangle$ with $b=1,2$ lives in a
four-dimensional Hilbert space and the entanglement of this bipartite
system can be described by the concurrence $C$ \cite{woo}. This
quantity is a measure of entanglement bounded in the range between
zero and one, with the maximum entanglement corresponding to unit
concurrence. On the other hand, for separable states, lacking any
entanglement, the concurrence is zero. 

Furthermore, due to the pure dephasing effect of the phonon scattering,
no transitions of the qubits are induced. Therefore, the phonon scattering
will transform an initial general superposition state
\begin{equation}
|\psi(0)\rangle=\sum_{m_{1},m_{2}=\pm\frac{\text{1}}{2}}\psi_{\{m_{1},m_{2}\}}|\{m_{1},m_{2}\}\rangle,
\end{equation}
into a non-pure statistical mixture of only those states that initially
already existed. Thus, during the time evolution the density operator
stays within the Hilbert sub space defined by the initial state. This
feature allows us to separately treat the two prominent cases of having
initially one or two {}``excitations'', respectively.

\subsection{Case of one excitation}

Assume the initial state of the two qubits to be of the form 
\begin{equation}
|\psi(0)\rangle=\psi_{\{\frac{1}{2},-\frac{1}{2}\}}|\{{\textstyle \frac{1}{2}},-{\textstyle \frac{1}{2}}\}\rangle+\psi_{\{-\frac{1}{2},\frac{1}{2}\}}|\{-{\textstyle \frac{1}{2}},{\textstyle \frac{1}{2}}\}\rangle.
\end{equation}
Since only one of the qubits is in its {}``excited'' state, this
superposition is usually denoted as the {}``one excitation'' case.
Choosing the basis vectors of the bipartite system as $\{|\{{\textstyle \frac{1}{2}},{\textstyle \frac{1}{2}}\}\rangle,|\{{\textstyle \frac{1}{2}},-{\textstyle \frac{1}{2}}\}\rangle,|\{-{\textstyle \frac{1}{2}},{\textstyle \frac{1}{2}}\}\rangle,|\{-{\textstyle \frac{1}{2}},-{\textstyle \frac{1}{2}}\}\rangle\}$,
following Eq. (\ref{eq:rho-offdiag2}) the time-dependent density
matrix can be written as 
\begin{equation}
\begin{tabular}{l}
 \ensuremath{\varrho_{{\rm S}}(t)=\left(\begin{array}{cccc}
0 & 0 & 0 & 0\\
0 & \varrho_{\{\frac{1}{2},-\frac{1}{2}\},\{\frac{1}{2},-\frac{1}{2}\}}(t) & \varrho_{\{\frac{1}{2},-\frac{1}{2}\},\{-\frac{1}{2},\frac{1}{2}\}}(t) & 0\\
0 & \varrho_{\{\frac{1}{2},-\frac{1}{2}\},\{-\frac{1}{2},\frac{1}{2}\}}^{\ast}(t) & \varrho_{\{-\frac{1}{2},\frac{1}{2}\},\{-\frac{1}{2},\frac{1}{2}\}}(t) & 0\\
0 & 0 & 0 & 0
\end{array}\right).}\end{tabular}\label{estado1}
\end{equation}
For this particular form of the density matrix, the concurrence simplifies
to 
\begin{equation}
C(t)=2\left|\varrho_{\{\frac{1}{2},-\frac{1}{2}\},\{-\frac{1}{2},\frac{1}{2}\}}(t)\right|.\label{eq:concurrence-1ex}
\end{equation}
Inserting the corresponding density matrix element from Eq. (\ref{eq:rho-offdiag2})
into Eq. (\ref{eq:concurrence-1ex}), the concurrence results as 
\begin{equation}
C(t)=2\sqrt{p(1-p)}\exp\left[-\int_{0}^{t}dt^{\prime}\Gamma_{\{\frac{1}{2},-\frac{1}{2}\}\{-\frac{1}{2},\frac{1}{2}\}}(t^{\prime})\right],\label{eq:concurrence-1ex-2}
\end{equation}
where $p=|\psi_{\{\frac{1}{2},-\frac{1}{2}\}}|^{2}$ is the initial
probability for the two qubits being in state $|\{{\textstyle \frac{1}{2}},-{\textstyle \frac{1}{2}}\}\rangle$.
The maximum initial concurrence is of course obtained for equal weights,
$p=1/2$, of the two constituent states. 

\begin{figure}
\begin{centering}
\includegraphics[width=1\columnwidth]{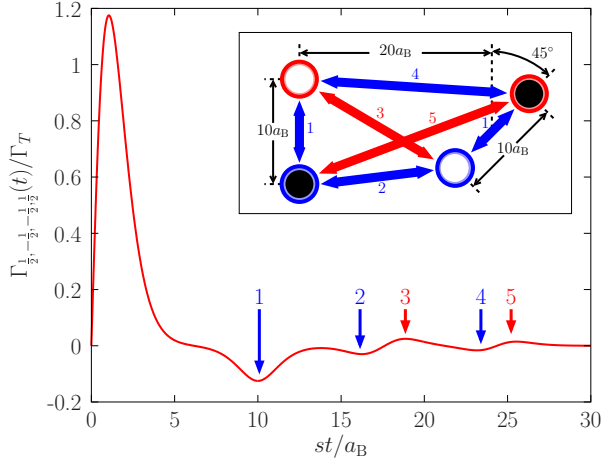}
\par\end{centering}

\caption{\label{fig:Dimensionless}Dimensionless disentanglement rate $\Gamma_{\{\frac{1}{2},-\frac{1}{2}\}\{-\frac{1}{2},\frac{1}{2}\}}(t)/\Gamma_{T}$
as a function of the dimensionless time $st/a_{{\rm B}}$ for the
case of a single excitation. The inset shows the geometrical configuration
of the donor sites: $d_{1}=d_{2}=10a_{{\rm B}}$, $|\vec{r}_{1}-\vec{r}_{2}|=20a_{{\rm B}}$,
$45\degree$ angle between qubit axes. The rounded lengths of the
inter-donor distances in units of $a_{{\rm B}}$ are: $l/a_{{\rm B}}=10$
(1), $16.53$ (2), $18.6$ (3), $23.6$ (4), $25.0$ (5).}

\end{figure}

From Eq. (\ref{eq:concurrence-1ex-2}) it becomes apparent that the
decoherence rate of the state of the qubits acts as disentanglement
rate. This rate is shown in Fig. \ref{fig:Dimensionless} for qubits
with $d=10a_{{\rm B}}$, an inter-qubit distance of $20a_{{\rm B}}$,
and a relative angle of $45\degree$, see inset of Fig. \ref{fig:Dimensionless}.
It shows a series of alternating maxima and minima at increasing times.
The principal positive maximum at the beginning occurs at $t\approx a_{{\rm B}}/s$,
which is the time needed by the phonon to travel within a donor site.
This is the main source of disentanglement. The times of the subsequent
extrema can be identified as the travel times between pairs of donor
sites, as indicated in the inset of Fig. \ref{fig:Dimensionless}.
Whereas the positive maxima destroy, the negative minima restore the
entanglement between the qubits. The mapping of which phonon path
between donor sites leads to positive or negative extrema in the disentanglement
rate can be established as follows: 

The concurrence is given by the modulus of the density matrix element
$\varrho_{\{\frac{1}{2},-\frac{1}{2}\},\{-\frac{1}{2},\frac{1}{2}\}}(t)$
that describes the time-dependent correlation between states $|\{-\frac{1}{2},\frac{1}{2}\}\rangle$
and $|\{\frac{1}{2},-\frac{1}{2}\}\rangle$. Correlations between
these states can only be created when a phonon travels between a donor
site occupied by one state to another donor site that is occupied
by the other state. The corresponding site occupations of each of
the states involved are indicated in the inset of Fig. \ref{fig:Dimensionless}
by black and white colors, respectively. The sites of each qubit are
marked by blue for $m_{b}=-\frac{1}{2}$ and red for $m_{b}=+\frac{1}{2}$.
With this color scheme, the creation of correlations mediated by sound
waves is produced via phonon travels between a black and a white donor
site. 

These phonon travels are: the passage within the individual qubits
for the distance $10a_{{\rm B}}$, that produces in Fig. \ref{fig:Dimensionless}
the negative minimum 1, and the passages between blue and red sites
at distances $\approx16.5a_{{\rm B}}$ and $\approx23.6a_{{\rm B}}$,
that produce the minima 2 and 4. All the other phonon passages produce
decorrelation and destroy the entanglement at distances $\approx18.6a_{{\rm B}}$
and $\approx25.0a_{{\rm B}}$, which generate the positive maxima
3 and 5 in Fig. \ref{fig:Dimensionless}.

\begin{figure}
\begin{centering}
\includegraphics[width=1\columnwidth]{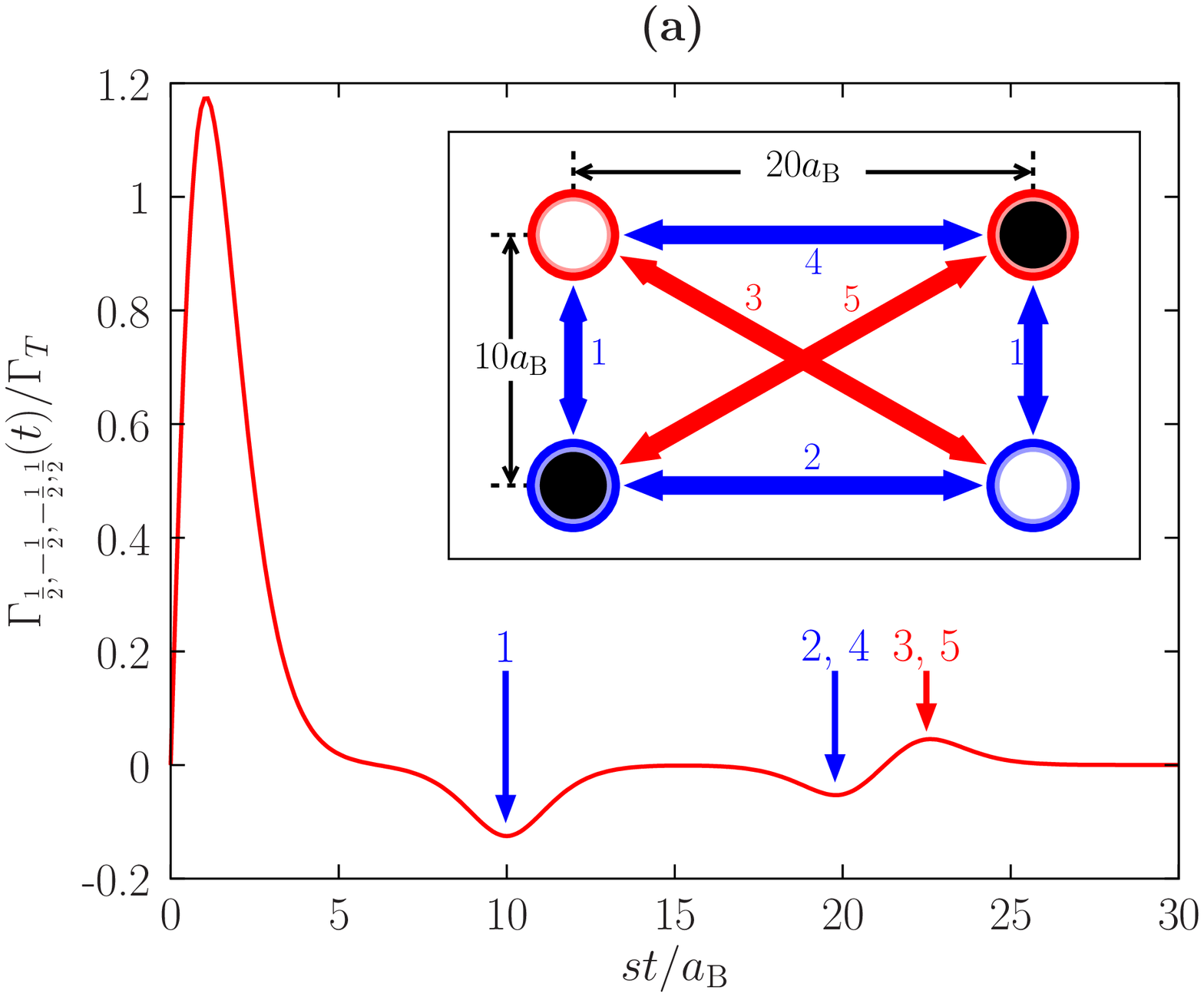}\vspace{5mm}

\par\end{centering}

\begin{centering}
\includegraphics[width=1\columnwidth]{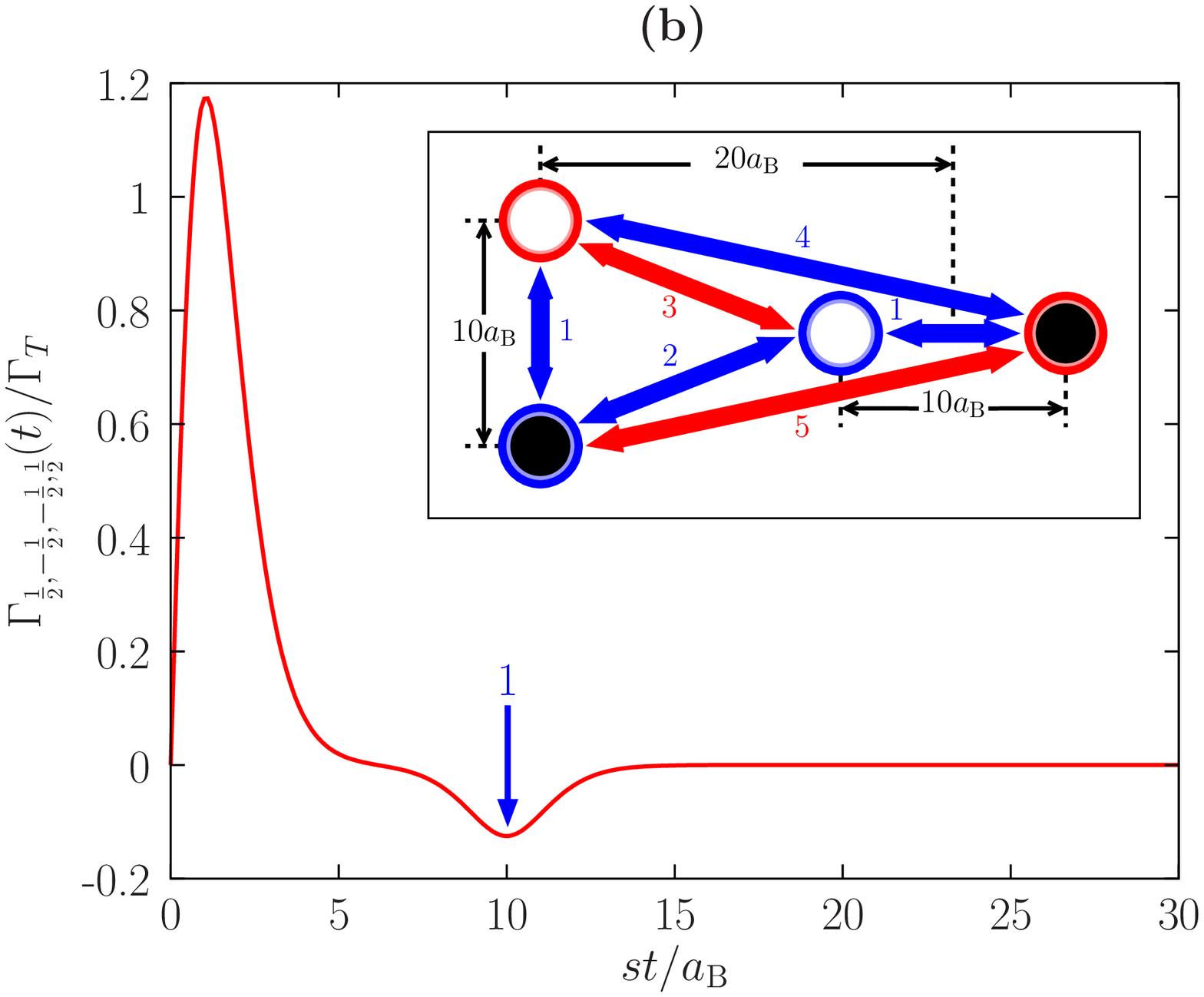}
\par\end{centering}

\caption{Dimensionless disentanglement rate $\Gamma_{\{\frac{1}{2},-\frac{1}{2}\}\{-\frac{1}{2},\frac{1}{2}\}}(t)/\Gamma_{T}$
as a function of the dimensionless time $st/a_{{\rm B}}$ for the
case of a single excitation. Same parameters as in Fig. 3, but for
collinear (a) and perpendicular (b) qubits.\label{fig:paralel}}

\end{figure}

We note that diminishing the $45\degree$ angle between the qubit
axes results in lengths 2 and 4, and lengths 3 and 5, respectively,
becoming progressively comparable. In the limiting case of two collinear
qubits, i.e. $0\degree$ angle, these pairs of lengths are identical
so that as a consequence the disentanglement rate shows only two positive
and two negative peaks, as shown in Fig. \ref{fig:paralel} (a). On
the other hand, as one approaches the limiting angle of $90\degree$,
i.e. the CNOT configuration \cite{hol,tsu,ste}, the extrema first
sparse and finally peaks 2 and 4 cancel peaks 3 and 5, respectively.
As a result, the disentanglement rate shows only the principal positive
peak at $t\approx a_{{\rm B}}/s$, see Fig. \ref{fig:paralel} (b).

\begin{figure}
\begin{centering}
\includegraphics[width=1\columnwidth]{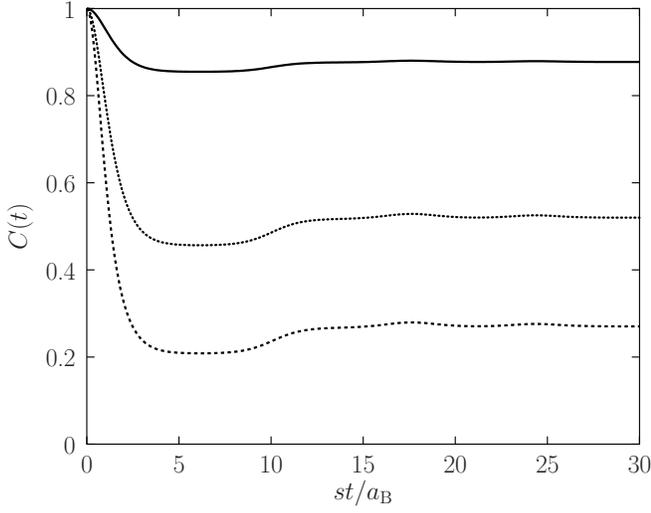}
\par\end{centering}

\caption{\label{fig:Concurrence-as-a}Concurrence as a function of the dimensionless
time $st/a_{{\rm B}}$ for the case of a single excitation for temperatures
$T/T_{{\rm B}}=0.01$ (solid curve), $0.05$ (dotted curve), $0.1$
(dashed curve). Other parameters same as in Fig. 3 with $p=1/2$.}

\end{figure}

The evolution of the concurrence for the case of $45\degree$ between
qubit axes is shown in Fig. \ref{fig:Concurrence-as-a}. It can be
seen that a stationary and non-vanishing value of the concurrence
is reached for large times. Moreover, the temperature dependence indicates
an only minor loss of entanglement at temperatures $T/T_{{\rm B}}<0.01$.
Given that for P impurities embedded in a Si substrate the characteristic
temperature is of the order of $T_{{\rm B}}\sim300\kelvin$, this
case corresponds to liquid He temperatures.

\subsection{Case of two excitations}

\begin{figure}[t]
\begin{centering}
\includegraphics[width=1\columnwidth]{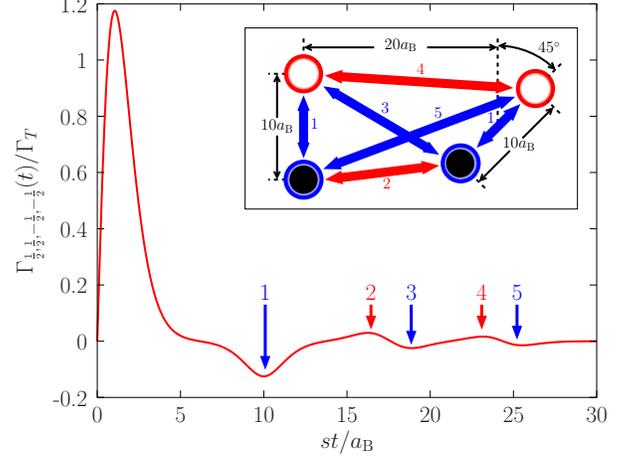}
\par\end{centering}

\centering{}\caption{\label{fig:Dimensionless-disentanglement-ra-10}Dimensionless disentanglement
rate $\Gamma_{\{\frac{1}{2},\frac{1}{2}\}\{-\frac{1}{2},-\frac{1}{2}\}}(t)/\Gamma_{T}$
as a function of the dimensionless time $st/a_{{\rm B}}$ for the
case of two excitations. The inset shows the geometrical configuration
of the donor sites: $d_{1}=d_{2}=10a_{{\rm B}}$, $|\vec{r}_{1}-\vec{r}_{2}|=20a_{{\rm B}}$,
$45\degree$ angle between qubit axes. The rounded lengths of the
inter-donor distances in units of $a_{{\rm B}}$ are: $l/a_{{\rm B}}=10$
(1), $16.53$ (2), $18.6$ (3), $23.6$ (4), $25.0$ (5).}
\end{figure}

The other prominent case is that of initial two excitations. This
case is described by the initial state being of the form 
\begin{equation}
|\psi(0)\rangle=\psi_{\{\frac{1}{2},\frac{1}{2}\}}|\{{\textstyle \frac{1}{2}},{\textstyle \frac{1}{2}}\}\rangle+\psi_{\{-\frac{1}{2},-\frac{1}{2}\}}|\{-{\textstyle \frac{1}{2}},-{\textstyle \frac{1}{2}}\}\rangle.
\end{equation}
In the same standard basis as before, the time-dependent density matrix
results then as 
\begin{equation}
\begin{tabular}{l}
 \ensuremath{\varrho(t)=\left(\begin{array}{cccc}
\varrho_{\{\frac{1}{2},\frac{1}{2}\},\{\frac{1}{2},\frac{1}{2}\}}(t) & 0 & 0 & \varrho_{\{\frac{1}{2},\frac{1}{2}\},\{-\frac{1}{2},-\frac{1}{2}\}}(t)\\
0 & 0 & 0 & 0\\
0 & 0 & 0 & 0\\
\varrho_{\{\frac{1}{2},\frac{1}{2}\},\{-\frac{1}{2},-\frac{1}{2}\}}^{\ast}(t) & 0 & 0 & \varrho_{\{-\frac{1}{2},-\frac{1}{2}\},\{-\frac{1}{2},-\frac{1}{2}\}}(t)
\end{array}\right).}\end{tabular}\label{estado2}
\end{equation}
Also for this case the concurrence simplifies to a simple expression,
given by 
\begin{equation}
C(t)=2\left|\varrho_{\{\frac{1}{2},\frac{1}{2}\},\{-\frac{1}{2},-\frac{1}{2}\}}(t)\right|,
\end{equation}
which, after insertion of Eq. (), becomes

\begin{equation}
C(t)=2\sqrt{p(1-p)}\exp\left[-\int_{0}^{t}dt^{\prime}\Gamma_{\{\frac{1}{2},\frac{1}{2}\},\{-\frac{1}{2},-\frac{1}{2}\}}(t^{\prime})\right],
\end{equation}
where now $p=|\psi_{\{\frac{1}{2},\frac{1}{2}\}}|^{2}$ is the probability
for the two qubits being {}``excited''. From the change of indices
it can be easily seen that, apart from the first positive and first
negative peak, the signs of the peaks of the disentanglement rate
are reversed as compared to the corresponding one excitation case,
see Fig. \ref{fig:Dimensionless-disentanglement-ra-10}.

\section{Summary and Outlook}

In this paper we considered the time evolution of entanglement in
donor-based charge qubits that is induced by off-resonant scattering
with acoustical phonons. We showed that this system can be solved
analytically and that a non-Markovian behavior emerges with negative
disentanglement rates, leading to non-monotonic disentanglement in
time. Moreover, for the cases of one and two initial excitations the
disentanglement rate is proportional to the decoherence rate of the
two-qubit state. In both cases the concurrence attains a stationary
and non-vanishing value at large times, which means that phonon scattering
does not completely destroy the entanglement of the initially prepared
two-qubit state. 

The choice of the geometry of the donor sites determines the features
of the concurrence as a time-dependent function. These features can
be understood by a simple kinetic interpretation of phonon travels
among the donor sites. In this work we focused on the cases of initially
one and two excitations. However, we believe that also particular
superpositions of both cases may be treated within this framework.
Furthermore, our model includes already the case of $N>2$ qubits,
where a trace over $N-2$ qubits would be required to obtain the entanglement
between a selected pair of qubits. This will be subject of future
work.
\begin{acknowledgments}
SW and FL acknowledge support by FONDECYT project no. 1095214. FL
acknowledges support from Financiamiento Basal project no. 0807.\end{acknowledgments}

\end{document}